\documentclass[twocolumn]{aastex63}

\usepackage{graphicx}
\usepackage{xspace}
\usepackage{mathtools}
\usepackage{amsmath}
\usepackage{gensymb}
\usepackage{lineno}

\def\hess{H.E.S.S.\xspace}
\def\j1702{HESS~J1702-420\xspace}
\def\ja1702{HESS~J1702-420A\xspace}
\def\jb1702{HESS~J1702-420B\xspace}

\hypersetup{linkcolor=red,citecolor=gray,filecolor=cyan,urlcolor=magenta}

\submitjournal{ApJ}

\shorttitle{Diffusion model for HESS~J1702-420}
\shortauthors{Aharonian et al}

\begin{document}

%\title{Relativistic diffusion model to explain spectral and morphological characteristics of \j1702 }
\title{On the nature of the energy-dependent morphology of the composite multi-TeV gamma-ray source \j1702}
%Relativistic protons diffusion as an explanation of energy-dependent morphology of \j1702 }
\author[0000-0003-1157-3915]{Felix Aharonian}\affiliation{Dublin Institute for Advanced Studies, School of Cosmic Physics, 31 Fitzwilliam Place, Dublin 2, Ireland}
\affiliation{Max-Planck-Institut f\"ur Kernphysik, Saupfercheckweg 1, 69117 Heidelberg, Germany}
\affiliation{Yerevan State University,  1 Alek Manukyan St, Yerevan 0025, Armenia}
\author[0000-0001-9689-2194]{Denys Malyshev}\affiliation{Institut f{\"u}r Astronomie und Astrophysik T{\"u}bingen, Universit{\"a}t T{\"u}bingen, Sand 1, D-72076 T{\"u}bingen, Germany}
\author[0000-0002-9735-3608]{Maria Chernyakova}\affiliation{School of Physical Sciences and Centre for Astrophysics \& Relativity, Dublin City University, Glasnevin, D09 W6Y4, Ireland}
\affiliation{Dublin Institute for Advanced Studies, School of Cosmic Physics, 31 Fitzwilliam Place, Dublin 2, Ireland}

\correspondingauthor{Felix Aharonian}
\email{Felix.Aharonian [at] mpi-hd.mpg.de}

\begin{abstract}
\j1702 is a multi-TeV gamma-ray source with an unusual energy-dependent morphology. The recent \hess observations suggest that the emission is well described by a combination of point-like \ja1702 (dominating at highest energies, $\gtrsim 30$~TeV ) and diffuse ($\sim 0.3^\circ$) \jb1702 (dominating below $\lesssim 5$\,TeV) sources with very hard ($\Gamma \sim 1.5$) and soft ($\Gamma \sim 2.6$) power-law spectra, respectively. Here we propose a model which postulates that the proton accelerator is located at the position of \ja1702 and is embedded into a dense molecular cloud that coincides with \jb1702.
In the proposed model, the VHE radiation of \j1702 is explained by the pion-decay emission from the continuously injected relativistic protons propagating through 
 a dense cloud. The energy-dependent morphology is defined by the diffusive nature of the low-energy protons propagation, 
transiting {\it sharply}  to (quasi) ballistic propagation at higher energies.  
Adopting strong energy dependence of the diffusion coefficient, $D \propto E^\beta$ with $\beta \geq 1$, we argue that \j1702 as the system of two gamma-ray sources is the result of the propagation effect. Protons injected by a single accelerator at the rate  $Q_0 \simeq    10^{38} \, (n_0/100 \, \rm cm^{-3})^{-1}\, (d/ \, 0.25\,kpc)^{-1} \rm  erg/s$  can reasonably reproduce the morphology and fluxes of two gamma-ray components.  
\end{abstract}


\section{Introduction}
\label{sec:intro}
\j1702 is a gamma-ray source discovered in the TeV band by the High Energy Spectroscopic System (\hess) during the first Galactic plane survey campaign \citep{hgps1}. Later, \citet{aharonian08} reported the extended morphology of the source and the first measurements of its spectral characteristics.

Recent \hess observations of \j1702 demonstrated that the morphology of the source is consistent with the superposition of emissions from a point-like central source \ja1702 and an extended source \jb1702 \citep{hess_j1702}. The point-like source is characterised by a power-law $\gamma$-ray spectrum with photon index $\Gamma_A=1.53\pm0.2$ extending without indication of steepening up to $\sim 100$~TeV. At low energies, below $\sim 5$~TeV, \ja1702 is outshone by \jb1702. The latter is characterised by a significantly softer spectrum with  $\Gamma_B=2.62\pm0.2$ and elliptical morphology with semi-axes of $0.32^\circ\pm 0.02^\circ_{stat}\pm 0.03^\circ_{syst}$ (major) and $0.20^\circ\pm 0.02^\circ_{stat}\pm 0.03^\circ_{syst}$(minor).

The origin of the gamma-ray emission from this source is unknown. Despite several dedicated deep X-ray observations with Suzaku~\citep{fujinaga11} and XMM-Newton~\citep{xmm22}, no clear counterparts were found for both the point-like \ja1702 and diffuse \jb1702 TeV sources. In the absence of clear spatially coincident counterpart sources at lower energies, several misplaced sources were invoked to explain the emission from \j1702. These include a cosmic ray diffusion from a nearby supernova remnant SNR G344.7-0.1 and a pulsar PSR J1702-412 (both $\sim 0.5^\circ$ away from the centroid of the TeV emission), see e.g. discussion in~\citet{hess_j1702}. 

In this paper, we propose a model that can explain the morphological and spectral characteristics of the emission coming from \j1702 region in a self-consistent way. We propose the point-like source \ja1702 to be a proton accelerator embedded into a dense molecular cloud. The diffuse source \jb1702 corresponds to the pion-decay emission from the continuously injected relativistic protons propagating through the cloud. The energy-dependent morphology of \j1702 (diffusive at $\geq  0.1$~TeV and point-like at $\gtrsim 10$~TeV) is explained by the diffusive nature of the low-energy protons propagation, which transits to almost rectilinear propagation of higher-energy protons. A similar scenario has been invoked to explain the spectrum of the  gamma-ray emission coming from the central region of our Galaxy at high and very high energies \citep{2005Ap&SS.300..255A, Chernyakova2011}

We describe the model in Section~\ref{sec:modelling}, discuss the results in Section~\ref{sec:results_discussion}, and summarize the conclusions in Section~\ref{sec:conclusion}.

\section{Modelling}
\label{sec:modelling}

The model postulates the presence of the high-energy proton accelerator embedded into a dense medium (molecular cloud).
To estimate the characteristic size and density of the ambient gas, we consider molecular HII clouds reported by ~\citet{Lau19} in the direction of \j1702. Several clouds were detected at distances from 0.25~kpc to $\sim 6$~kpc and with characteristic number densities $10^2-10^3$~cm$^{-3}$. Below we will discuss the closest ($d=0.25$~kpc) cloud from \citet{Lau19} characterized by density $n_0=180 \, \rm cm^{-3}$ within $0.32^\circ$. It corresponds to the cloud's radius $R=1.4$~pc and a rather modest mass of about 100  $M_\odot$. Later, we will discuss how the derived results could be rescaled for more distant and heavy clouds.   

The stationary distribution function $f(r,\mu)$ of relativistic protons injected by a point-like source and propagating through the ambient medium is given by~\cite{prosekin15}:
\begin{align}
\label{eq:distr_function}
& f(r,\mu) = \frac{Q}{8\pi^2 c Z}\left( \frac{1}{r^2} + \frac{c}{rD} \right)  \ \exp\left(-\frac{3D(1-\mu)}{rc}\right) ; \\  \nonumber
& Z(x) = \frac{x}{3}\left(1-e^{-6/x}\right).
\end{align}
Here $r$ is the radial coordinate; the source of the relativistic protons with the energy-dependent injection rate $Q(E)$ is assumed to be located at $r=0$; $\mu=\cos\theta$ is the cosine of the angle between proton propagation and radial direction. The transport of protons is described by the energy-dependent diffusion coefficient $D=D(E)$.  Eq.~(\ref{eq:distr_function}) provides the radial distribution of relativistic protons in both diffusion and ballistic regimes, including the transition between these two propagation modes.  

Below we parametrize the injection power of relativistic protons $Q(E)$  and the diffusion coefficient $D(E)$  as
\begin{align}
\label{eq:parameterisation}
& Q(E_p) \equiv dN_p/dE_p = N_0\cdot \left(E_p/1\,\mbox{TeV}\right)^{-\alpha}e^{-E_p/E_{cut}} \\ \nonumber
& D(E_p) = D_0\cdot\left(E_p/1\,\mbox{TeV}\right)^{\beta} \, ,
\end{align}
with the total energetics  in accelerated  protons $Q_0 = \int\limits_{m_p}^{\infty} E_p\cdot Q(E_p)dE_p$. 

The relativistic protons, during their propagation through the ambient gas,  emit gamma rays in $pp$ collisions. The gamma-ray spectra were calculated with the help of \texttt{naima} v.0.9.1~\cite{naima} python module, which for the pion decay emission channel implements the parametrisation from \cite{kafexhiu14}.

%%%%%%%%%%%%%%%%%%%%%%%%%%%%%%%%%%%%%% Fig 1
\begin{figure*}
\includegraphics[width=0.49\linewidth]{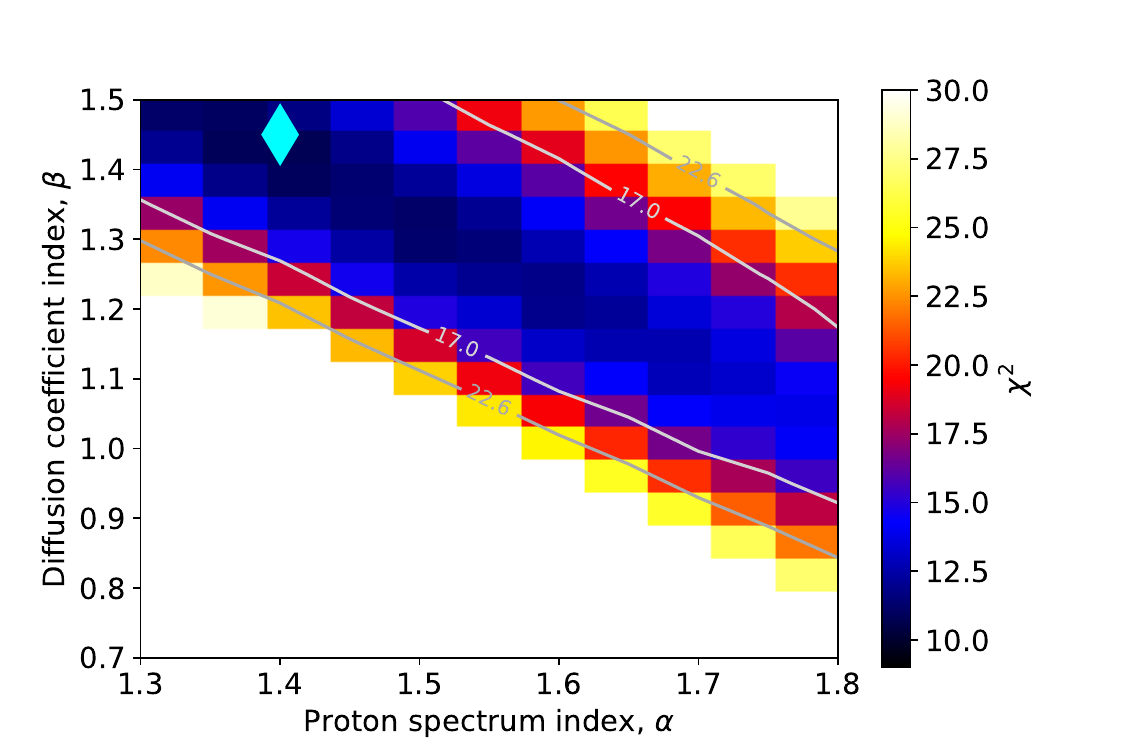}
\includegraphics[width=0.49\linewidth]{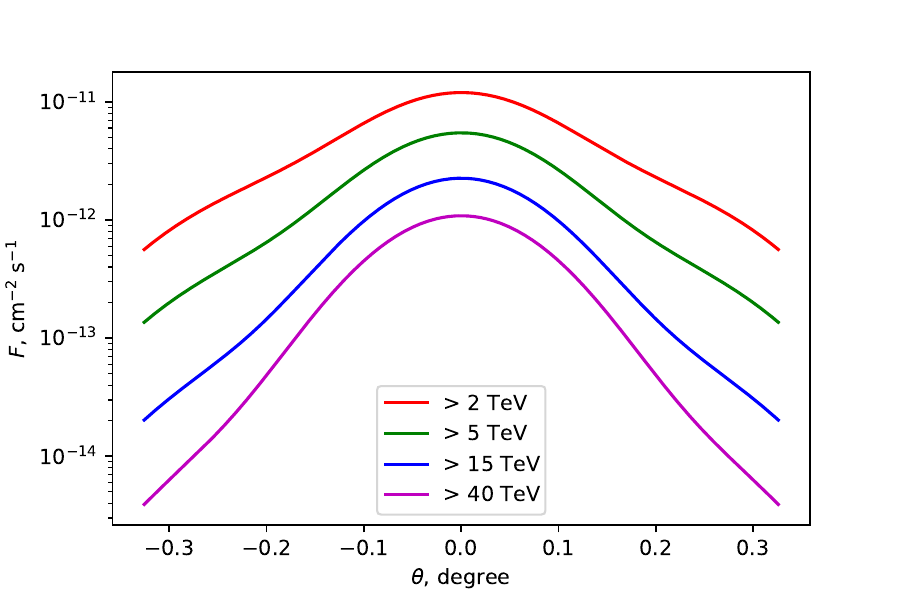}
\caption{\textit{Left:} $1\sigma$ and $2\sigma$ contours on the parameter set  ($\beta,\alpha$)  ($D\propto E^{\beta}$, $dN_p/dE_p\propto E_p^{-\alpha}$). The cyan diamond corresponds to the overall best-fit combination of $\beta$ and $\alpha$. \textit{ Right:} Brightness profiles at different energies as seen above-specified energies smoothed with \hess PSF (adopted as Gaussian with dispersion $0.07^\circ$), similar to the procedure used} in~\cite{hess_j1702} 
\label{fig:contours}
\end{figure*}
%%%%%%%%%%%%%%%%%%%%%%%%%%%%%%%%%%%%%%

The results for the intensity and morphology of the produced emission were obtained by integrating the produced gamma-ray emission over the production region. 
Based on the derived intensity profile, we extracted the model spectrum of the diffuse source from the $0.15^\circ-0.3^\circ$ annulus. The spectrum of the point source was extracted from a central $0.15^\circ$-radius circle which roughly corresponds to the  95\% \hess PSF (assumed to be a Gaussian with $0.07^\circ$ dispersion) containment. We note that within the performed modelling, the diffuse source presents a natural background for the point-like source. In order to minimize the effects of this background, we additionally subtracted the scaled (according to the extraction area) diffuse source spectrum from the point source spectrum.

Following the described procedure, we derived the spectra of the point-like and diffuse sources for a set of model parameters -- diffusion coefficient index $\beta$, diffusion coefficient normalisation $D_0$ and the power-law index of the injected proton spectrum,  $\alpha$. In the absence of firm indications for the high-energy cutoff in the observed spectra of point-like and diffuse gamma-ray sources up to $E \sim 100$~TeV,  we set $E_{\rm cut}=1$~PeV, implying that we deal with a {\it Proton PeVatron}.   The observed gamma-ray spectrum of \ja1702 proposes a  hard power-law spectrum  of injected protons with   $\alpha <2$. Thus  the total energy in protons is determined essentially by the  upper limit, while the lower limit of integration doesn't have a significant impact on the proton injection power: $Q_{0} = \int\limits_{\mbox{\tiny{1 TeV}}}^{\mbox{\tiny{1 PeV}}} E_p Q(E_p)dE_p$.   Because of the lack of evidence of a high energy cutoff in the gamma-ray spectrum, this estimate should be considered as a lower limit. 

Within the proposed model, the fluxes of both \ja1702 and \jb1702  are proportional to the product of the proton injection power  $Q_{0}$ and the target gas density $n_0$. Thus, $Q_{0}n_0$ can be derived through the joint fitting of the model fluxes of two sources to the observed ones.

\section{Results and Discussion}
\label{sec:results_discussion}
\subsection{Modelling results}
\label{sec:modelling_results}
\begin{figure*}
\includegraphics[width=0.49\linewidth]{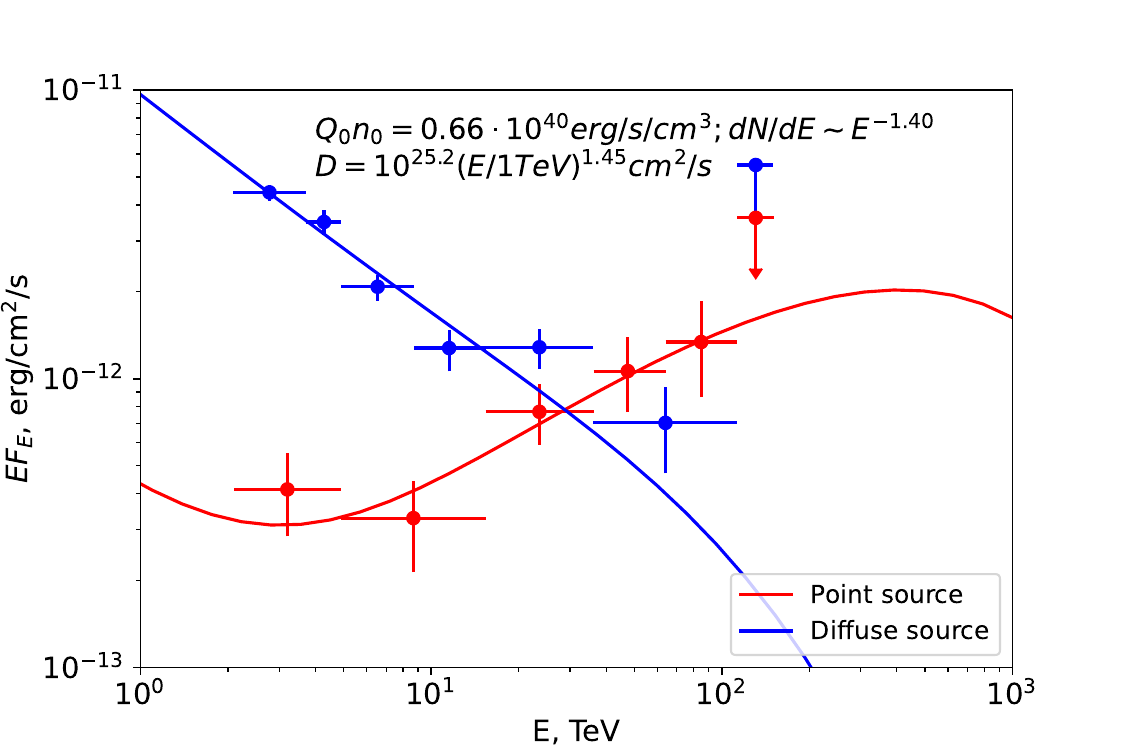}
\includegraphics[width=0.49\linewidth]{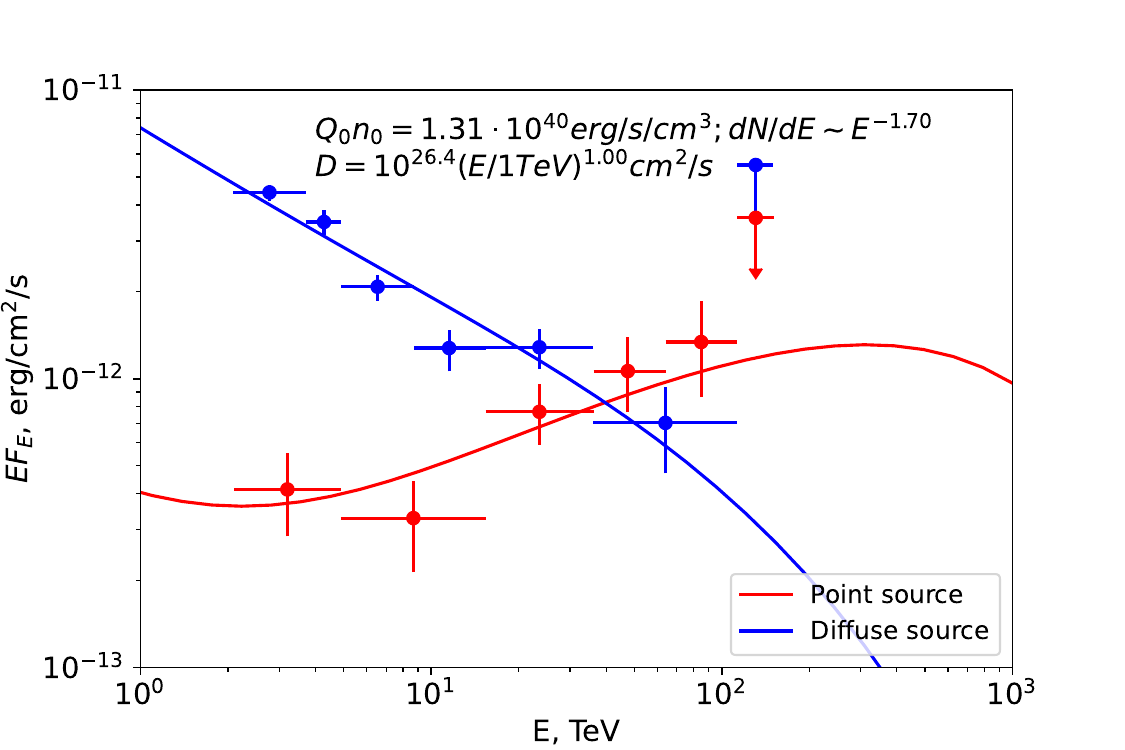}
\caption{The observed and calculated spectral energy distributions (SEDs) of the point-like and diffuse sources for two sets of model parameters $(\beta,\alpha)$.   \textit{Left:}  the best-fit parameters corresponding to the cyan diamond point shown in Fig~\ref{fig:contours}:   $\beta=1.45$  and $\alpha=1.4$. \ \textit{Right:} The model spectra are calculated for not the best but still an acceptable set of parameters with   Bohm-type diffusion coefficient index,  $\beta=1$, and the proton's spectral index   $\alpha=1.7$.  The corresponding values of the product $Q_0 n_0$ are  $0.66 \times 10^{40} \, \rm erg/cm^3 s$  and $1.31 \times 10^{40} \, \rm erg/cm^3 s$, respectively.  
%The selected values of parameters are consistent with data at a better than $2\sigma$ level.  
}
\label{fig:best_fit_VHE_spectra}  
\end{figure*}

 The gamma-ray brightness profiles at different energies, convolved with the \hess PSF, are shown in Fig.~\ref{fig:contours} (right panel). They show a tendency of reduction of the source's angular size from low to high energies. This trend is clearly seen in  Fig.~\ref{fig:best_fit_VHE_spectra}, which demonstrates a gradual transformation of the flux dominance by the diffuse source \jb1702 at low energies to the flux dominance of the point-like source \ja1702 at highest energies.   

For the parameter set ($\beta$, $\alpha$), the $\chi^2$ range of the joint model fit of the point-like and diffuse source spectra is shown in the left panel of Fig.~\ref{fig:contours}. The areas within grey contours are consistent with the data at $2\sigma$ (light grey, inner contour) and $3\sigma$ (dark grey, outer contour) levels. The numbers on the contours indicate the corresponding $\chi^2$ values (best-fit $\chi^2_{0}=10.8$).  The cyan diamond point corresponds to the formal best-fit value of all parameters found during the fit. The best-fit parameters provide a good fit to the data for  the diffusion coefficient $D = 1.6\cdot 10^{25} (E/1\mbox{TeV})^{1.45}$~cm$^2$/s, the product  $Q_{0}n_0=0.66\cdot 10^{40}$~erg/s/cm$^3$ and the power-law index of proton's spectrum $\alpha=1.4$.  However, the strong correlation of $\beta$ and $\alpha$ parameters (see  Fig.~\ref{fig:contours})  doesn't allow us to derive the  $\beta$  and $\alpha$ indices separately.  Indeed,  the spectral index of relativistic protons modulated by diffusion is $\approx \alpha+\beta$~\citep{aharonian_atoyan96}.  Correspondingly,  the gamma-ray photon index,  which roughly mimics the slope of the proton spectrum  (due to the almost energy-independence of  $pp$ interaction cross-section), contains information about the sum of two indices, $\alpha+\beta$.   We should also note that the results rely only on statistical uncertainties of data which are at the level of 6-30\% of measured fluxes. The systematic uncertainty of 10-20\% typical for \hess data~\citep{hess_crab} can further broaden the allowed parameter space shown in Fig.~\ref{fig:contours}.   
For example,  one cannot exclude the combination of  ($\beta=1, \alpha=1.7$), which gives an acceptable fit to the data;  see the right panel of Fig.~\ref{fig:best_fit_VHE_spectra}.  Nevertheless, despite these uncertainties,  the calculations show that the diffusion coefficient should have a  sharp energy dependence, namely   $\beta \geq 1$.

%%%%%%%%%%%%%%%%%%%%%%%%%%%%%%%%%%%%%%
%%%%%%%%%%%%%%%%%%%%%%%%%%%%%%%%%%%%%% 

%The best-fit value of the diffusion coefficient index is $\beta=1.45$. 
%Generally, the index of the diffusion coefficient  is considered to be small,  $\beta \leq 1$.  For example, in the case of the most effective diffusion in the so-called Bohm regime, $\beta=1$. However, the recent studies of particle diffusion in highly turbulent environments not only allow but, in some cases, give  preference to sharp energy-dependence of the diffusion coefficient with $\beta \geq 1$ \citep{Reichherzer2020, Reichherzer2022}.  {\bf Note that $\beta \geq 1$ characterizes the tendency of transition from diffusive to ballistic  propagation regime}. 

Generally, the index of the diffusion coefficient in a variety of standard astronomical environments, e.g. in the interstellar medium or in supernova remnants,  is small,   $\beta \leq 1$.  For example, in the  Kolmogorov and  Kraichnan turbulence modes, $\beta=1/3 $ and $1/2$, respectively, achieving $\beta=$ 1 in the Bohm diffusion regime.  In this regard, the sharp energy dependence of the diffusion coefficient, which is a strongly preferred option in our model, can be considered somewhat unusual and suspicious. However, the recent studies of particle diffusion in highly turbulent environments not only allow but, in some cases, give preference to sharp energy-dependence of the diffusion coefficient with $\beta \geq 1$ \citep{Giacinti,Reichherzer2020, Reichherzer2022}.  The discussion of this non-trivial theoretical issue is outside the scope of this paper; here, we limit it by noticing that  $\beta \geq 1$ characterizes the tendency of faster transition from diffusive to ballistic propagation regime and thus can be considered a natural consequence of our phenomenological model.  We also note that for both parameter sets shown Fig.~\ref{fig:best_fit_VHE_spectra},  the diffusion coefficient at low (TeV) energies is by orders of magnitude smaller than in the interstellar medium (ISM) ($\sim 10^{29-30}$~cm$^2$/s at 1\,TeV, see e.g.~\citet{strong07} for a review).  This is another example that in different gamma-ray source populations, e.g. in Pulsar Halos~\citep{2017Sci...358..911A} and  Stellar Clusters~\citep{Aharonian2019},  CR diffusion may proceed in a very slow regime.  At UHE energies,  thanks to the strong energy dependence, the diffusion coefficient quickly recovers,  although it still remains below the characteristic for the ISM level.   But it appears to be sufficient to deviate at these energies from the nominal diffusion,  namely to move, on the pc scales of the cloud,  (quasi) ballistically. As a result, despite the large angular size ($\approx 0.3^\circ$) of the cloud where gamma-rays are produced, the apparent size of the gamma-ray image at multi-TeV energies is less than the HESS PSF of about $0.1^\circ$. At lower energies, because of the diffusive character of propagation, the angular size of the gamma-ray image coincides with the cloud's angular size.

\begin{figure}
\includegraphics[width=\linewidth]{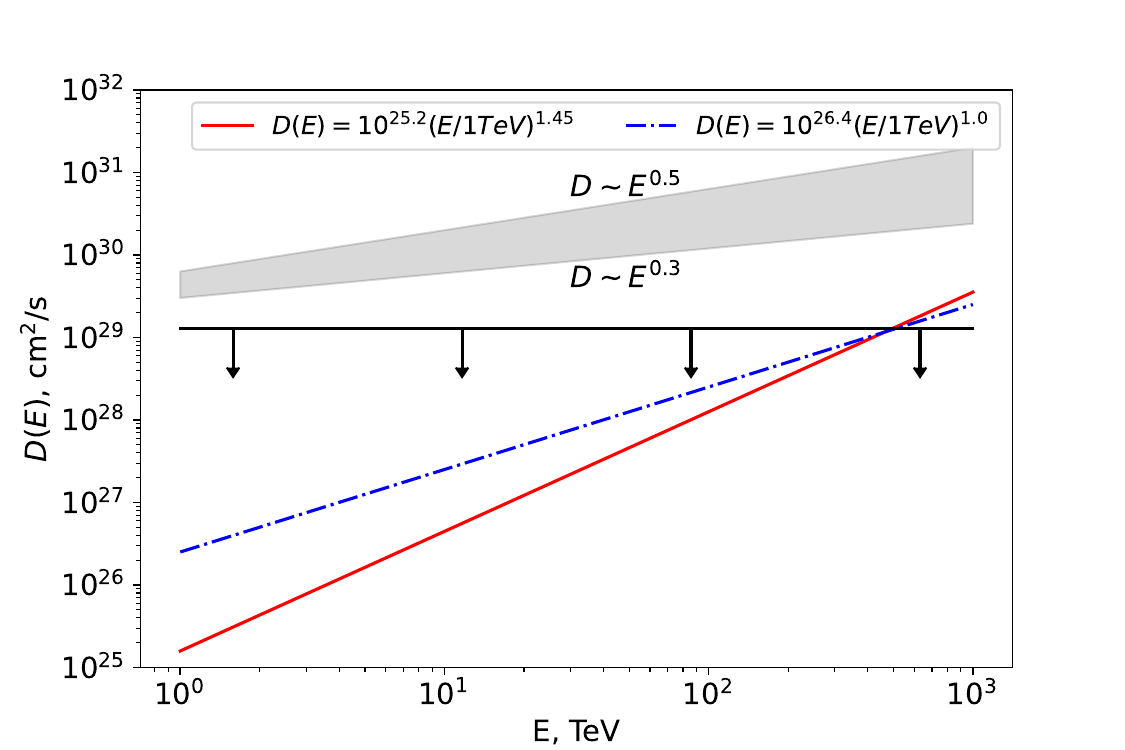}
\caption{The energy dependencies of two diffusion coefficients used to describe HESS data (see Fig.~\ref{fig:best_fit_VHE_spectra}). The value of the diffusion coefficient corresponding to the transition from diffusive to (quasi)ballistic propagation is shown with the horizontal black line. The shaded region corresponds to the range of the diffusion coefficient adopted for galactic cosmic rays ~\citep{strong07,vladimirov12}. }
\label{fig:diffusion_coefficients}
\end{figure}

The energy dependencies of two diffusion coefficients used in  Fig.~\ref{fig:best_fit_VHE_spectra} are shown in Fig.~\ref{fig:diffusion_coefficients} with solid (red) and dot-dashed (blue) lines. The horizontal black line presents the margin of applicability of the diffusive propagation regime defined $D_{max}=Rc$, as it follows from Eq.~(\ref{eq:distr_function}). Above that line,  the propagation proceeds in the ballistic regime. For comparison, in Fig.~\ref{fig:diffusion_coefficients}, we show the range of the diffusion coefficient commonly adopted for galactic cosmic rays ~\citep{strong07,vladimirov12}.

%The obtained fit is acceptable for a broad correlated range of $\beta$ and $\alpha$ parameters. For example,  the best-fit for  $\beta=1$ is acceptable and consistent with the data at a better than $2\sigma$ level, see right panel of Fig.~\ref{fig:best_fit_VHE_spectra} for the following parameters: $D = 3.2 \times 10^{26} (E/1\mbox{TeV})^{1}$~cm$^2$/s, $Q_{0}n_0=0.82\cdot 10^{40}$~erg/s/cm$^3$, $\alpha=1.7$.
%Note that $\beta=1$ describes  the particle transport  in the Bohm diffusion  limit. However, the absolute value of the required diffusion coefficient  exceeds the  Bohm diffusion coefficient by 1.5 orders of magnitude. Still, it is much less than the standard coefficients for cosmic rays ($\sim 10^{31}$~cm$^2$/s at 1\,TeV, see e.g.~\citet{strong07} for a review).

% The brightness profiles of the TeV emission (convolved with \hess PSF) are shown in Fig.~\ref{fig:contours}, right panel. The brightness profiles demonstrate a trend of the decreasing of characteristic size from low to high energies. The same trend -- a gradual transformation from a diffuse low-energy source \jb1702 to a point-like high-energy source \ja1702. The best-fit spectra for cases of the formal best-fit parameters (left panel) and a Bohm-like diffusion (right panel) are shown in Fig.~\ref{fig:best_fit_VHE_spectra}.

\subsection{Estimates for an arbitrary cloud}
\label{sec:quick_and_dirty_estimations}
The proposed model successfully explains the spectra and morphology of \j1702 VHE sources, assuming the distance to the source of 0.25~kpc.  For the given angular scales, the distance to the source determines the geometrical size of the source, which is one of the principal model parameters for the description of the CR transport, including the transition from the diffusive to ballistic propagation regimes.  The impact of the ambient gas density on the results is simpler; the gamma-ray flux is proportional to $n_0$ and, consequently, the proton injection power  $Q_0 \propto 1/n_0$.   If the gamma-ray source coincides with the cloud reported in \cite{Lau19} at a distance $d \approx 0.25 \, \rm kpc$ with density  $n_0 \approx  180 \, \rm cm^{-3}$,  the required injection power  $Q_0 \approx  3.6 \times 10^{37} \, \rm erg/s$  for $\beta=1.4$ and twice larger for  $\beta = 1$.

Since one cannot exclude the association of the gamma-ray source with other clouds reported by \cite{Lau19}, 
below, we briefly discuss how the main parameters could be rescaled for a cloud located at an arbitrary distance.

%\subsubsection{Diffusion coefficient}
Diffusion softens the proton's energy distribution and, consequently, the resulting gamma-ray spectrum by $\beta$. Thus the index $\beta$ of the diffusion coefficient could be estimated from the difference of spectral slopes of \ja1702 ($\Gamma_A=1.53$) and \jb1702 ($\Gamma_B=2.62$) sources: $\beta \simeq \Gamma_B-\Gamma_A \sim  1$.
The transition from the diffusive to (quasi) rectilinear propagation of protons occurs at energy $E_{\rm tr}$ defined from the condition $R^2/\left(2D(E_{\rm tr})\right) \sim R/c$; see Eq.~(\ref{eq:distr_function}):

%On the other hand, the relativistic protons are moving through the cloud with a bulk velocity $v=2D(E)/R$ and protons with energies above

\begin{equation}
\label{eq:Ec}  
\frac{E_{\rm tr}}{1\mbox{TeV}}\lesssim\left ( \frac{cR}{2D_0} \right )^{1/\beta} \simeq \left ( 2100 \frac{L} {0.25\mbox{\,kpc}} \frac {10^{26}\mbox{cm$^2$s$^{-1}$}}{D_0} \right)^{1/\beta}
\end{equation}
 
Protons with higher energies propagate almost ballistically, therefore above $E_{\rm tr}$ the proton spectrum is not modified. Correspondingly, we should expect a noticeable change in the gamma-ray spectrum around $E_{\rm tr,\gamma}\sim 0.1  E_{\rm tr}$.
 For the same reason, we should expect different gamma-ray images at high and low energies, namely,  an extended source below $E_{\rm tr,\gamma}$, and a point-like source above $E_{\rm tr,\gamma}$. These predictions describe quite well the observed energy-dependent morphology of \j1702 and allow us to estimate the absolute value of the diffusion coefficient $D_0$ and $E_{\rm tr,\gamma}$  -- the energy of transition from a point-like to diffuse morphology. 

%%%%%%%%%%%%%%%%%%%%%%%%%%%%%%%%%%%%%%
%%%%%%%%%%%%%%%%%%%%%%%%%%%%%%%%%%%%%%
 
For the diffusion coefficients derived from the numerical modelling,  one gets $E_{\rm tr}\sim 700$~TeV. 
For a hard proton spectrum  ($\alpha\lesssim 2$), the ratio of energies of the primary proton and secondary ($\pi^0$-decay) photon is about 20  \citep{Kelner2006, Celli2020}, thus the transition between point-like and diffuse morphologies should occur occurs at $E_{\rm tr,\gamma}\sim 35$~TeV, in a good agreement with numerical calculations shown in Fig.~\ref{fig:best_fit_VHE_spectra}.

 \subsubsection{Proton's injection rate}

The flux, angular size and distance to HESS J1702-420A determine the required proton injection.
The efficiency of conversion of the energy of CR protons to gamma-rays in inelastic $pp$  interactions is determined by the ratio of the confinement time  $t_{\rm esc}$  of protons inside the emitter to their radiative cooling time through the production and decay of  $\pi^0$- mesons: $$ t_{pp \to \pi^0} \simeq 5  \times 10^{15} \left(n_0/1 \mbox{\,cm}^{-3}\right)^{-1} \mbox{s.}$$
In general, the energy-dependent confinement time is determined by the diffusion coefficient, but for the estimates of gamma-ray flux from the point-like source,  we should use  $t_{\rm esc} \approx R/c$  given that the latter is the result of radiation by protons moving (quasi) ballistically inside the source of the size $R\approx \theta d$, where  $\theta \approx 0.32^\circ$  is the angular size of the emitter (the diffuse source coinciding with the cloud in the Proton PeVatron is embedded), $d$ is the distance to the system.  Then, the flux expected from the point-like source is:
 \begin{align}
 \label{Fph}
 & F_{\gamma}= \frac{Q_0}{4\pi d^2} \frac {t_{esc}}{t_{pp\rightarrow \pi^0}} \simeq \\ \nonumber
 & \simeq  4\cdot 10^{-12}\frac{Q_{0}n_0}{10^{40}\mbox{erg/s/cm}^3}\left(\frac{d}{0.25\mbox{\,kpc}}\right)^{-1}\left(\frac{\theta_{\rm }}{0.32^\circ}\right) \frac{\mbox{erg}}{\mbox{cm}^2s},
 \end{align}
 
 %Here, we assume that about 10\% of protons' energy is released in photons.
  Note that in this equation, the gamma-ray flux scales with distance as $\propto d^{-1}$. % , therefore the injection rate of protons 
% \begin{equation}
% Q_0 \simeq 10^{38} \left(\frac{d}{1 \rm kpc}\right) \, \left(\frac{n_0}{100 \, \rm cm^{-3}}\right)^{-1} \ \left(\frac{F_{\rm obs}}{10^{-12} \, \frac{\rm erg}{\rm cm^2 s}}\right) \, \frac{\rm erg}{\rm s}     
% \end{equation}
 
 %Equation~(\ref{Fph}) gives an estimate of the $Q_{0} n_0 \simeq 10^{40}$~erg/s/cm$^3$ assuming  $F_{\gamma}=10^{-12}$~erg/cm$^2$/s in agreement  with values reported in Sec.~\ref{sec:modelling_results}.  

\subsection{X-ray Emission from Secondary Electrons}

In $pp$ collisions, about the same number of gamma rays and electrons are produced.  Propagating through the ambient magnetic field (in the cloud or in the ISM),  the secondary electrons radiate potentially detectable synchrotron emission. For the characteristic magnetic  field,   $B_0=10\,\mu$G~\citep{cox05,jansson_farrar}, the electrons with energy $E_e$  are  cooled on  timescale 
\begin{align}
\label{eq:synch_tcool}
& t_{\rm cool} = 6.6\cdot 10^2 \left(\frac{B}{10\,\mu\mbox{G}}\right)^{-2}\left(\frac{E_e}{100\,\mbox{TeV}}\right)^{-1}\,\mbox{yr} \, .
\end{align}
The Larmor radius $R_L$ and the characteristic distance of the propagating electrons  $s$ are
%in the case of Bohm diffusion are
\begin{align}
\label{eq:RL_sbohm}
& R_L \simeq  10^{-2} \left(\frac{B}{10\,\mu\mbox{G}}\right)^{-1}  \left(\frac{E_e}{100\,\mbox{TeV}}\right)\,\mbox{pc} \\ \nonumber
& s(t) = \sqrt{D t_{cool}}   \simeq  \\ \nonumber
& 4.7 \sqrt{\frac{D_{0}}{10^{26}\mbox{cm}^2/\mbox{s}}} \left(\frac{E_e}{1\,\mbox{TeV}} \right)^{(\beta-1)/2}  \left(\frac{B}{10\,\mu\mbox{G}}\right)^{-1} \mbox{pc} \\ \nonumber \
%& s_{bohm}(t) = \sqrt{D_{bohm}t} = \sqrt{\frac{1}{3}cR_Lt} = \\ \nonumber
%& = 3 \left(\frac{B}{10\,\mu\mbox{G}}\right)^{-1/2}  \left(\frac{E_e}{100\,\mbox{TeV}}\right)^{1/2} \left(\frac{t}{1000\,\mbox{yr}}\right)^{1/2}\,\mbox{pc}
\end{align}

An electron of energy $E_e$ emits synchrotron emission at  
\begin{align}
& \varepsilon_s = 5 \frac{B}{10\,\mu\mbox{G}}  \left(\frac{E_e}{100\,\mbox{TeV}}\right)^2 \mbox{keV,}
\end{align}
thus the propagation distance can be written as
\begin{align}
& s(t) =  4.7 \sqrt{\frac{D_{0}}{10^{26}\mbox{cm}^2/\mbox{s}}} \left(\frac{\varepsilon_s}{0.5\,\mbox{eV}} \right)^{\frac{\beta-1}{4}}  \left(\frac{B}{10\,\mu\mbox{G}}\right)^{\frac{-\beta-3}{4}} \mbox{pc} 
\end{align}

In the case of  $\beta=1$ and assuming $D_0=10^{26.5}$ cm$^2$/s,  this equation  reduces to 
\begin{align}
& s \simeq  8 \left(\frac{B}{10\,\mu\mbox{G}}\right)^{-1} \mbox{pc}  \, . 
\label{eq:sbohm}
\end{align}
Thus, for the distance to the cloud $d=0.25$~kpc, the angular size of the X-ray image, $\theta = s/d \sim 2^\circ$, significantly exceeds the cloud's angular size.  
%To estimate the angular size of  X-ray emission for a cloud located at a larger distance, one should keep in mind that as it follows from the equation~(\ref{eq:Ec}), the required best-fit $D_0$ value is proportional to the size of the source, which in turn is proportional to the source distance, as the angular size of the source is fixed. Thus the characteristic path of the propagation of the electron is proportional to the square root of the source distance. For example, for the cloud located at 3.5~kpc, as it was assumed in the \cite{hess_j1702}, the angular size of the X-ray source would reduce to $\sim 0.5^\circ$, compatible with XMM-Newton field of view.
%
As it follows from Eq.(\ref{eq:sbohm}), only in the case of a much stronger magnetic field, namely, $B \gtrsim 50\,\mu$G, the X-ray image could be smaller than the size of the molecular cloud. To calculate the synchrotron flux of the secondary electrons produced, we used the aafragpy v.1.12 package~\citep{aafrag} for the electron production in $pp$ interactions, and the naima v.0.9.1~\cite{naima}
python module for the synchrotron radiation  in a random magnetic 
field.

%%%%%%%%%%%%%%%%%%%%%%%%%%%%%%%%%%%%  Fig.4
\begin{figure}
\includegraphics[width=\linewidth]{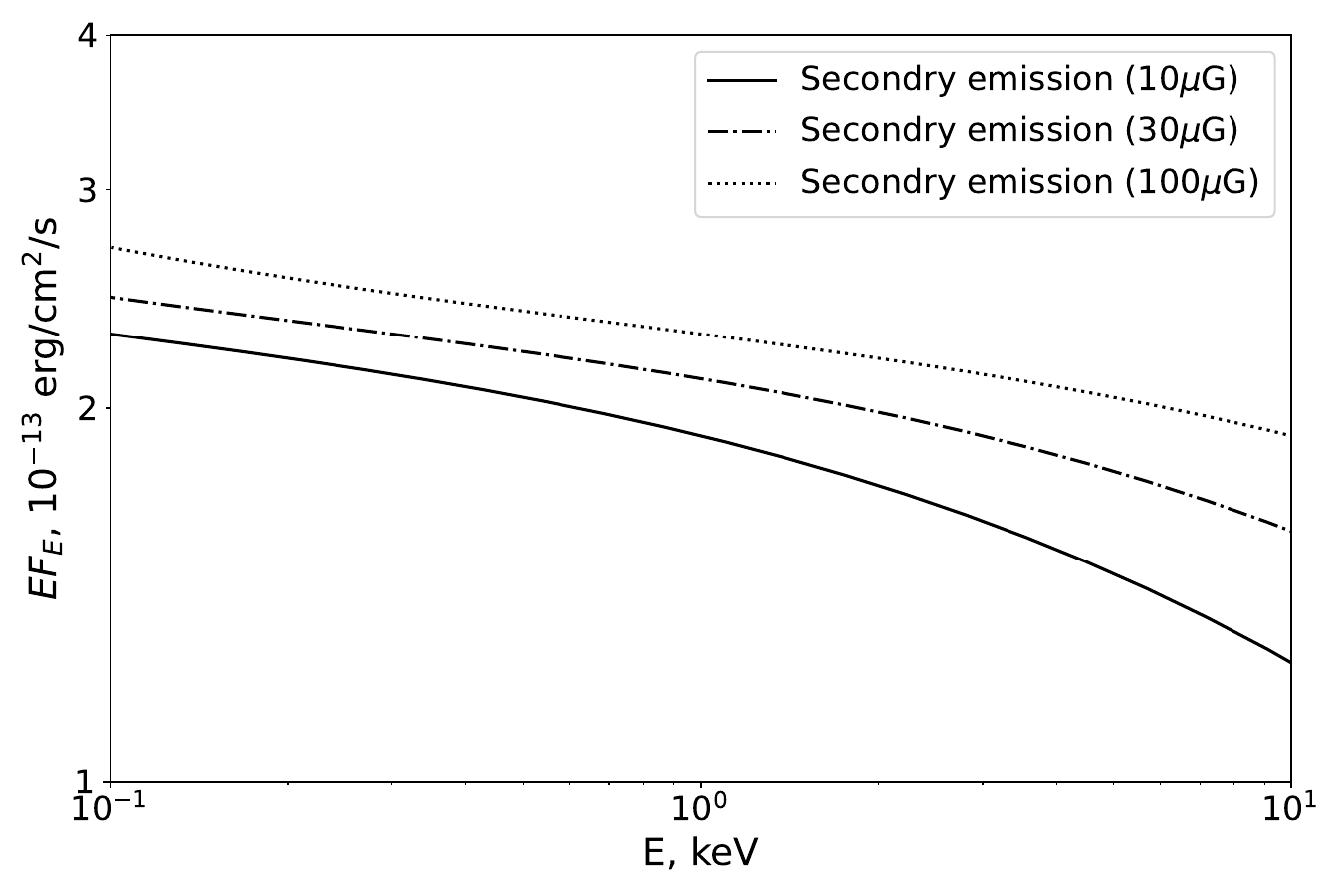}
\caption{ Modelled X-ray spectra of the secondary synchrotron emission
integrated over the X-ray emitting region for different values of the magnetic field.}
\label{fig:xray_model}
\end{figure}
%%%%%%%%%%%%%%%%%%%%%%%%%%%%%%%%%%%%%%

The fluxes of the synchrotron emission of the secondary electrons integrated over the entire  X-ray source for three values of the magnetic field are shown in Fig~\ref{fig:xray_model}. The absolute fluxes are obtained from the normalization to flux of 
the $\pi^0$-decay $\gamma$-rays.  The X-ray spectrum is hard, with a photon index slightly exceeding $2$ and the flux $\sim 2-2.5 \times 10^{-13}$~erg/cm$^2$/s at 1~keV.  Since the entire energy of the secondary electrons is emitted via synchrotron radiation, the X-ray flux only slightly depends on the strength of the magnetic field. The increase of the latter results only in the shift of the energy of the synchrotron photons proportional to $B$.

\cite{xmm22} carried out a deep X-ray observation of \j1702 with XMM-Newton resulted in a non-detection of a counterpart of a point-like source \ja1702. This result is consistent with our model, as $\sim 100$~TeV secondary electrons producing keV emission propagate through the medium in the diffusive regime (as follows e.g. from Eq.~(\ref{eq:synch_tcool}) and Eq.~(\ref{eq:RL_sbohm})). The consequent emission corresponds to the extended morphology of $\sim 2^\circ$ angular size as discussed above, which significantly exceeds the field of view (FoV) of XMM-Newton, making problematic the detection of the extended emission with pointing observations. The larger-scale mosaic XMM-Newton observations or surveys with broad FoV missions, e.g. eROSITA potentially are able to detect the predicted X-ray emission. 
Note that in the soft gamma-ray (MeV) band, the source becomes point-like as the corresponding secondary electrons propagate in the ballistic regime. Formally, this makes detection easier. However, the expected flux level $\sim 10^{-13}$~erg/cm$^2$/s at 1~MeV is still by two orders of magnitude below the sensitivities of the current and future missions.

\section{Summary}
\label{sec:conclusion}
%%%%%%%%%%%%%%%%%%%%%%%%%%%%%%%%%%%%%%%%%%%%%%%%%%%%%%%%%%%
\j1702 is a TeV gamma-ray source of particular interest because of its hard energy spectrum extending up to 100 TeV and its peculiar energy-dependent morphology.  

This paper proposes a model that addresses the spatial distribution and energy spectra of the extended and point-like components of \j1702 originated by a single accelerator (a proton PeVatron) embedded in a dense gas cloud.   The observed emission from \j1702 is explained by $\pi^0$-decay gamma rays arising from the interactions of relativistic protons continuously injected into and propagating through the cloud. The energy-dependent gamma-ray morphology is caused by the diffusive nature of the propagation of low-energy protons, which transits to an almost ballistic propagation regime at the highest energies. For a reasonable set of model parameters, both the energy spectrum and morphology can be well described by the diffusion coefficient, which is essentially suppressed at low energies in comparison to the interstellar medium ($D_0 \sim 10^{26}$~cm$^2$/s at 1 TeV) but with a strong energy-dependence ($\beta \gtrsim 1$) that results in the propagation of highest energy protons ($E \geq 100$~TeV) in ballistic regime. The detected fluxes of gamma rays require a powerful proton accelerator with an injection rate at the level of  $Q_0 \sim 10^{38}(n_0/100 \mbox{cm}^{-3})^{-1}$~erg/s.

We argue that the proposed scenario can be typical for a broad class of multi-TeV gamma ray sources.

\section*{Acknowledgements}
We thank the anonymous referee for his/her thoughtful comments, which 
helped us to improve the manuscript. The work of DM was supported by DLR through grant 50OR2104 and by DFG through grant MA 7807/2-1. The authors acknowledge support by the state of Baden-W\"urttemberg through~bwHPC.

 %%%%%%%%%%%%%%%%%%%%%%%%%%%%%%%%%%%%%%%%%%%%%%%%%%%%%%%%%%%%%%%

%The best-fit model parameters correspond to $D = 10^{25.2} (E/1\mbox{TeV})^{1.45}$~cm$^2$/s, $Q_{100}n_0=0.16\cdot 10^{40}$~erg/s/cm$^3$, $\alpha=1.35$. 

%$0.05^\circ$-radius and the hard spectrum with index $\sim -2$. The flux of the source is at the level detectable by modern all-sky X-ray observatories, e.g. eROSITA which is able to detect extended sources with the characteristic flux of $few\times 10^{-13}$~erg/cm$^2$/s~\cite{erosita_science_book}.

%Non-detection of the source in the X-ray band could indicate the non-stationary processes in the source, e.g. that the source started it's VHE activity less than synchrotron cooling time $t_{cool}\sim 1000$~yr ago.

% Bibliography and bibfile
\def\aj{AJ}%
          % Astronomical Journal
\def\actaa{Acta Astron.}%
          % Acta Astronomica
\def\araa{ARA\&A}%
          % Annual Review of Astron and Astrophys
\def\apj{ApJ}%
          % Astrophysical Journal
\def\apjl{ApJ}%
          % Astrophysical Journal, Letters
\def\apjs{ApJS}%
          % Astrophysical Journal, Supplement
\def\ao{Appl.~Opt.}%
          % Applied Optics
\def\apss{Ap\&SS}%
          % Astrophysics and Space Science
\def\aap{A\&A}%
          % Astronomy and Astrophysics
\def\aapr{A\&A~Rev.}%
          % Astronomy and Astrophysics Reviews
\def\aaps{A\&AS}%
          % Astronomy and Astrophysics, Supplement
\def\azh{AZh}%
          % Astronomicheskii Zhurnal
\def\baas{BAAS}%
          % Bulletin of the AAS
\def\bac{Bull. astr. Inst. Czechosl.}%
          % Bulletin of the Astronomical Institutes of Czechoslovakia
\def\caa{Chinese Astron. Astrophys.}%
          % Chinese Astronomy and Astrophysics
\def\cjaa{Chinese J. Astron. Astrophys.}%
          % Chinese Journal of Astronomy and Astrophysics
\def\icarus{Icarus}%
          % Icarus
\def\jcap{J. Cosmology Astropart. Phys.}%
          % Journal of Cosmology and Astroparticle Physics
\def\jrasc{JRASC}%
          % Journal of the RAS of Canada
\def\mnras{MNRAS}%
          % Monthly Notices of the RAS
\def\memras{MmRAS}%
          % Memoirs of the RAS
\def\na{New A}%
          % New Astronomy
\def\nar{New A Rev.}%
          % New Astronomy Review
\def\pasa{PASA}%
          % Publications of the Astron. Soc. of Australia
\def\pra{Phys.~Rev.~A}%
          % Physical Review A: General Physics
\def\prb{Phys.~Rev.~B}%
          % Physical Review B: Solid State
\def\prc{Phys.~Rev.~C}%
          % Physical Review C
\def\prd{Phys.~Rev.~D}%
          % Physical Review D
\def\pre{Phys.~Rev.~E}%
          % Physical Review E
\def\prl{Phys.~Rev.~Lett.}%
          % Physical Review Letters
\def\pasp{PASP}%
          % Publications of the ASP
\def\pasj{PASJ}%
          % Publications of the ASJ
\def\qjras{QJRAS}%
          % Quarterly Journal of the RAS
\def\rmxaa{Rev. Mexicana Astron. Astrofis.}%
          % Revista Mexicana de Astronomia y Astrofisica
\def\skytel{S\&T}%
          % Sky and Telescope
\def\solphys{Sol.~Phys.}%
          % Solar Physics
\def\sovast{Soviet~Ast.}%
          % Soviet Astronomy
\def\ssr{Space~Sci.~Rev.}%
          % Space Science Reviews
\def\zap{ZAp}%
          % Zeitschrift fuer Astrophysik
\def\nat{Nature}%
          % Nature
\def\iaucirc{IAU~Circ.}%
          % IAU Cirulars
\def\aplett{Astrophys.~Lett.}%
          % Astrophysics Letters
\def\apspr{Astrophys.~Space~Phys.~Res.}%
          % Astrophysics Space Physics Research
\def\bain{Bull.~Astron.~Inst.~Netherlands}%
          % Bulletin Astronomical Institute of the Netherlands
\def\fcp{Fund.~Cosmic~Phys.}%
          % Fundamental Cosmic Physics
\def\gca{Geochim.~Cosmochim.~Acta}%
          % Geochimica Cosmochimica Acta
\def\grl{Geophys.~Res.~Lett.}%
          % Geophysics Research Letters
\def\jcp{J.~Chem.~Phys.}%
          % Journal of Chemical Physics
\def\jgr{J.~Geophys.~Res.}%
          % Journal of Geophysics Research
\def\jqsrt{J.~Quant.~Spec.~Radiat.~Transf.}%
          % Journal of Quantitiative Spectroscopy and Radiative Trasfer
\def\memsai{Mem.~Soc.~Astron.~Italiana}%
          % Mem. Societa Astronomica Italiana
\def\nphysa{Nucl.~Phys.~A}%
          % Nuclear Physics A
\def\physrep{Phys.~Rep.}%
          % Physics Reports
\def\physscr{Phys.~Scr}%
          % Physica Scripta
\def\planss{Planet.~Space~Sci.}%
          % Planetary Space Science
\def\procspie{Proc.~SPIE}%
          % Proceedings of the SPIE
\let\astap=\aap
\let\apjlett=\apjl
\let\apjsupp=\apjs
\let\applopt=\ao
\bibliographystyle{aasjournal}
\bibliography{bibliography}

\begin{thebibliography}{}
\expandafter\ifx\csname natexlab\endcsname\relax\def\natexlab#1{#1}\fi
\providecommand{\url}[1]{\href{#1}{#1}}
\providecommand{\dodoi}[1]{doi:~\href{http://doi.org/#1}{\nolinkurl{#1}}}
\providecommand{\doeprint}[1]{\href{http://ascl.net/#1}{\nolinkurl{http://ascl.net/#1}}}
\providecommand{\doarXiv}[1]{\href{https://arxiv.org/abs/#1}{\nolinkurl{https://arxiv.org/abs/#1}}}

\bibitem[{{Abdalla} {et~al.}(2021){Abdalla}, {Aharonian}, {Ait Benkhali},
  {Ang{\"u}ner}, {Arcaro}, {Armand}, {Armstrong}, {Ashkar}, {Backes},
  {Baghmanyan}, {Barbosa Martins}, {Barnacka}, {Barnard}, {Becherini}, {Berge},
  {Bernl{\"o}hr}, {Bi}, {B{\"o}ttcher}, {Boisson}, {Bolmont}, {de Bony de
  Lavergne}, {Breuhaus}, {Brun}, {Brun}, {Bryan}, {B{\"u}chele}, {Bulik},
  {Bylund}, {Caroff}, {Carosi}, {Casanova}, {Chand}, {Chandra}, {Chen},
  {Cotter}, {Cury{\l}o}, {Damascene Mbarubucyeye}, {Davids}, {Davies}, {Deil},
  {Devin}, {Dirson}, {Djannati-Ata{\"\i}}, {Dmytriiev}, {Donath}, {Doroshenko},
  {Dreyer}, {Duffy}, {Dyks}, {Egberts}, {Eichhorn}, {Einecke}, {Emery},
  {Ernenwein}, {Feijen}, {Fegan}, {Fiasson}, {Fichet de Clairfontaine},
  {Fontaine}, {Funk}, {F{\"u}{\ss}ling}, {Gabici}, {Gallant}, {Giavitto},
  {Giunti}, {Glawion}, {Glicenstein}, {Grondin}, {Hahn}, {Haupt}, {Hermann},
  {Hinton}, {Hofmann}, {Hoischen}, {Holch}, {Holler}, {H{\"o}rbe}, {Horns},
  {Huber}, {Jamrozy}, {Jankowsky}, {Jankowsky}, {Jardin-Blicq}, {Joshi},
  {Jung-Richardt}, {Kasai}, {Kastendieck}, {Katarzy{\'n}ski}, {Katz},
  {Khangulyan}, {Kh{\'e}lifi}, {Klepser}, {Klu{\'z}niak}, {Komin}, {Konno},
  {Kosack}, {Kostunin}, {Kreter}, {Lamanna}, {Lemi{\`e}re}, {Lemoine-Goumard},
  {Lenain}, {Leuschner}, {Levy}, {Lohse}, {Lypova}, {Mackey}, {Majumdar},
  {Malyshev}, {Malyshev}, {Marandon}, {Marchegiani}, {Marcowith}, {Mares},
  {Mart{\'\i}-Devesa}, {Marx}, {Maurin}, {Meintjes}, {Meyer}, {Mitchell},
  {Moderski}, {Mohrmann}, {Montanari}, {Moore}, {Morris}, {Moulin}, {Muller},
  {Murach}, {Nakashima}, {Nayerhoda}, {de Naurois}, {Ndiyavala}, {Niemiec},
  {Oakes}, {O'Brien}, {Odaka}, {Ohm}, {Olivera-Nieto}, {de Ona Wilhelmi},
  {Ostrowski}, {Panny}, {Panter}, {Parsons}, {Peron}, {Peyaud}, {Piel}, {Pita},
  {Poireau}, {Priyana Noel}, {Prokhorov}, {Prokoph}, {P{\"u}hlhofer}, {Punch},
  {Quirrenbach}, {Raab}, {Rauth}, {Reichherzer}, {Reimer}, {Reimer}, {Remy},
  {Renaud}, {Rieger}, {Rinchiuso}, {Romoli}, {Rowell}, {Rudak}, {Ruiz-Velasco},
  {Sahakian}, {Sailer}, {Salzmann}, {Sanchez}, {Santangelo}, {Sasaki},
  {Scalici}, {Sch{\"a}fer}, {Sch{\"u}ssler}, {Schutte}, {Schwanke},
  {Seglar-Arroyo}, {Senniappan}, {Seyffert}, {Shafi}, {Shapopi},
  {Shiningayamwe}, {Simoni}, {Sinha}, {Sol}, {Specovius}, {Spencer},
  {Spir-Jacob}, {Stawarz}, {Sun}, {Steenkamp}, {Stegmann}, {Steinmassl},
  {Steppa}, {Takahashi}, {Tavernier}, {Taylor}, {Terrier}, {Thiersen},
  {Tiziani}, {Tluczykont}, {Tomankova}, {Trichard}, {Tsirou}, {Tuffs},
  {Uchiyama}, {van der Walt}, {van Eldik}, {van Rensburg}, {van Soelen},
  {Vasileiadis}, {Veh}, {Venter}, {Vincent}, {Vink}, {V{\"o}lk}, {Wadiasingh},
  {Wagner}, {Watson}, {Werner}, {White}, {Wierzcholska}, {Wun Wong},
  {Yusafzai}, {Zacharias}, {Zanin}, {Zargaryan}, {Zdziarski}, {Zech}, {Zhu},
  {Zorn}, {Zouari}, \& {{\.Z}ywucka}}]{hess_j1702}
{Abdalla}, H., {Aharonian}, F., {Ait Benkhali}, F., {et~al.} 2021, \aap, 653,
  A152, \dodoi{10.1051/0004-6361/202140962}

\bibitem[{{Abeysekara} {et~al.}(2017){Abeysekara}, {Albert}, {Alfaro},
  {Alvarez}, {{\'A}lvarez}, {Arceo}, {Arteaga-Vel{\'a}zquez}, {Avila Rojas},
  {Ayala Solares}, {Barber}, {Bautista-Elivar}, {Becerril}, {Belmont-Moreno},
  {BenZvi}, {Berley}, {Bernal}, {Braun}, {Brisbois}, {Caballero-Mora},
  {Capistr{\'a}n}, {Carrami{\~n}ana}, {Casanova}, {Castillo}, {Cotti},
  {Cotzomi}, {Couti{\~n}o de Le{\'o}n}, {De Le{\'o}n}, {De la Fuente},
  {Dingus}, {DuVernois}, {D{\'\i}az-V{\'e}lez}, {Ellsworth}, {Engel},
  {Enr{\'\i}quez-Rivera}, {Fiorino}, {Fraija}, {Garc{\'\i}a-Gonz{\'a}lez},
  {Garfias}, {Gerhardt}, {Gonz{\'a}lez Mu{\~n}oz}, {Gonz{\'a}lez}, {Goodman},
  {Hampel-Arias}, {Harding}, {Hern{\'a}ndez}, {Hern{\'a}ndez-Almada}, {Hinton},
  {Hona}, {Hui}, {H{\"u}ntemeyer}, {Iriarte}, {Jardin-Blicq}, {Joshi},
  {Kaufmann}, {Kieda}, {Lara}, {Lauer}, {Lee}, {Lennarz}, {Vargas},
  {Linnemann}, {Longinotti}, {Luis Raya}, {Luna-Garc{\'\i}a}, {L{\'o}pez-Coto},
  {Malone}, {Marinelli}, {Martinez}, {Martinez-Castellanos},
  {Mart{\'\i}nez-Castro}, {Mart{\'\i}nez-Huerta}, {Matthews},
  {Miranda-Romagnoli}, {Moreno}, {Mostaf{\'a}}, {Nellen}, {Newbold}, {Nisa},
  {Noriega-Papaqui}, {Pelayo}, {Pretz}, {P{\'e}rez-P{\'e}rez}, {Ren}, {Rho},
  {Rivi{\`e}re}, {Rosa-Gonz{\'a}lez}, {Rosenberg}, {Ruiz-Velasco}, {Salazar},
  {Salesa Greus}, {Sandoval}, {Schneider}, {Schoorlemmer}, {Sinnis}, {Smith},
  {Springer}, {Surajbali}, {Taboada}, {Tibolla}, {Tollefson}, {Torres},
  {Ukwatta}, {Vianello}, {Weisgarber}, {Westerhoff}, {Wisher}, {Wood},
  {Yapici}, {Yodh}, {Younk}, {Zepeda}, {Zhou}, {Guo}, {Hahn}, {Li}, \&
  {Zhang}}]{2017Sci...358..911A}
{Abeysekara}, A.~U., {Albert}, A., {Alfaro}, R., {et~al.} 2017, Science, 358,
  911, \dodoi{10.1126/science.aan4880}

\bibitem[{{Aharonian} \& {Neronov}(2005)}]{2005Ap&SS.300..255A}
{Aharonian}, F., \& {Neronov}, A. 2005, \apss, 300, 255,
  \dodoi{10.1007/s10509-005-1209-4}

\bibitem[{{Aharonian} {et~al.}(2019){Aharonian}, {Yang}, \& {de O{\~n}a
  Wilhelmi}}]{Aharonian2019}
{Aharonian}, F., {Yang}, R., \& {de O{\~n}a Wilhelmi}, E. 2019, Nature
  Astronomy, 3, 561, \dodoi{10.1038/s41550-019-0724-0}

\bibitem[{{Aharonian} {et~al.}(2006{\natexlab{a}}){Aharonian}, {Akhperjanian},
  {Bazer-Bachi}, {Beilicke}, {Benbow}, {Berge}, {Bernl{\"o}hr}, {Boisson},
  {Bolz}, {Borrel}, {Braun}, {Breitling}, {Brown}, {Chadwick}, {Chounet},
  {Cornils}, {Costamante}, {Degrange}, {Dickinson}, {Djannati-Ata{\"\i}},
  {Drury}, {Dubus}, {Emmanoulopoulos}, {Espigat}, {Feinstein}, {Fontaine},
  {Fuchs}, {Funk}, {Gallant}, {Giebels}, {Gillessen}, {Glicenstein}, {Goret},
  {Hadjichristidis}, {Hauser}, {Heinzelmann}, {Henri}, {Hermann}, {Hinton},
  {Hofmann}, {Holleran}, {Horns}, {Jacholkowska}, {de Jager}, {Kh{\'e}lifi},
  {Komin}, {Konopelko}, {Latham}, {Le Gallou}, {Lemi{\`e}re},
  {Lemoine-Goumard}, {Leroy}, {Lohse}, {Martin}, {Martineau-Huynh},
  {Marcowith}, {Masterson}, {McComb}, {de Naurois}, {Nolan}, {Noutsos},
  {Orford}, {Osborne}, {Ouchrif}, {Panter}, {Pelletier}, {Pita},
  {P{\"u}hlhofer}, {Punch}, {Raubenheimer}, {Raue}, {Raux}, {Rayner}, {Reimer},
  {Reimer}, {Ripken}, {Rob}, {Rolland}, {Rowell}, {Sahakian}, {Saug{\'e}},
  {Schlenker}, {Schlickeiser}, {Schuster}, {Schwanke}, {Siewert}, {Sol},
  {Spangler}, {Steenkamp}, {Stegmann}, {Tavernet}, {Terrier}, {Th{\'e}oret},
  {Tluczykont}, {Vasileiadis}, {Venter}, {Vincent}, {V{\"o}lk}, \&
  {Wagner}}]{hgps1}
{Aharonian}, F., {Akhperjanian}, A.~G., {Bazer-Bachi}, A.~R., {et~al.}
  2006{\natexlab{a}}, \apj, 636, 777, \dodoi{10.1086/498013}

\bibitem[{{Aharonian} {et~al.}(2006{\natexlab{b}}){Aharonian}, {Akhperjanian},
  {Bazer-Bachi}, {Beilicke}, {Benbow}, {Berge}, {Bernl{\"o}hr}, {Boisson},
  {Bolz}, {Borrel}, {Braun}, {Breitling}, {Brown}, {B{\"u}hler},
  {B{\"u}sching}, {Carrigan}, {Chadwick}, {Chounet}, {Cornils}, {Costamante},
  {Degrange}, {Dickinson}, {Djannati-Ata{\"\i}}, {O'C. Drury}, {Dubus},
  {Egberts}, {Emmanoulopoulos}, {Espigat}, {Feinstein}, {Ferrero}, {Fiasson},
  {Fontaine}, {Funk}, {Funk}, {Gallant}, {Giebels}, {Glicenstein}, {Goret},
  {Hadjichristidis}, {Hauser}, {Hauser}, {Heinzelmann}, {Henri}, {Hermann},
  {Hinton}, {Hofmann}, {Holleran}, {Horns}, {Jacholkowska}, {de Jager},
  {Kh{\'e}lifi}, {Komin}, {Konopelko}, {Kosack}, {Latham}, {Le Gallou},
  {Lemi{\`e}re}, {Lemoine-Goumard}, {Lohse}, {Martin}, {Martineau-Huynh},
  {Marcowith}, {Masterson}, {McComb}, {de Naurois}, {Nedbal}, {Nolan},
  {Noutsos}, {Orford}, {Osborne}, {Ouchrif}, {Panter}, {Pelletier}, {Pita},
  {P{\"u}hlhofer}, {Punch}, {Raubenheimer}, {Raue}, {Rayner}, {Reimer},
  {Reimer}, {Ripken}, {Rob}, {Rolland}, {Rowell}, {Sahakian}, {Saug{\'e}},
  {Schlenker}, {Schlickeiser}, {Schwanke}, {Sol}, {Spangler}, {Spanier},
  {Steenkamp}, {Stegmann}, {Superina}, {Tavernet}, {Terrier}, {Th{\'e}oret},
  {Tluczykont}, {van Eldik}, {Vasileiadis}, {Venter}, {Vincent}, {V{\"o}lk},
  {Wagner}, \& {Ward}}]{hess_crab}
---. 2006{\natexlab{b}}, \aap, 457, 899, \dodoi{10.1051/0004-6361:20065351}

\bibitem[{{Aharonian} {et~al.}(2008){Aharonian}, {Akhperjanian}, {Barres de
  Almeida}, {Bazer-Bachi}, {Behera}, {Beilicke}, {Benbow}, {Bernl{\"o}hr},
  {Boisson}, {Bolz}, {Borrel}, {Braun}, {Brion}, {Brown}, {B{\"u}hler},
  {Bulik}, {B{\"u}sching}, {Boutelier}, {Carrigan}, {Chadwick}, {Chounet},
  {Clapson}, {Coignet}, {Cornils}, {Costamante}, {Dalton}, {Degrange},
  {Dickinson}, {Djannati-Ata{\"\i}}, {Domainko}, {Drury}, {Dubois}, {Dubus},
  {Dyks}, {Egberts}, {Emmanoulopoulos}, {Espigat}, {Farnier}, {Feinstein},
  {Fiasson}, {F{\"o}rster}, {Fontaine}, {Funk}, {F{\"u}{\ss}ling}, {Gallant},
  {Giebels}, {Glicenstein}, {Gl{\"u}ck}, {Goret}, {Hadjichristidis}, {Hauser},
  {Hauser}, {Heinzelmann}, {Henri}, {Hermann}, {Hinton}, {Hoffmann}, {Hofmann},
  {Holleran}, {Hoppe}, {Horns}, {Jacholkowska}, {de Jager}, {Jung},
  {Katarzy{\'n}ski}, {Kendziorra}, {Kerschhaggl}, {Kh{\'e}lifi}, {Keogh},
  {Komin}, {Kosack}, {Lamanna}, {Latham}, {Lemi{\`e}re}, {Lemoine-Goumard},
  {Lenain}, {Lohse}, {Martin}, {Martineau-Huynh}, {Marcowith}, {Masterson},
  {Maurin}, {Maurin}, {McComb}, {Moderski}, {Moulin}, {de Naurois}, {Nedbal},
  {Nolan}, {Ohm}, {Olive}, {de O{\~n}a Wilhelmi}, {Orford}, {Osborne},
  {Ostrowski}, {Panter}, {Pedaletti}, {Pelletier}, {Petrucci}, {Pita},
  {P{\"u}hlhofer}, {Punch}, {Ranchon}, {Raubenheimer}, {Raue}, {Rayner},
  {Renaud}, {Ripken}, {Rob}, {Rolland}, {Rosier-Lees}, {Rowell}, {Rudak},
  {Ruppel}, {Sahakian}, {Santangelo}, {Schlickeiser}, {Sch{\"o}ck},
  {Schr{\"o}der}, {Schwanke}, {Schwarzburg}, {Schwemmer}, {Shalchi}, {Sol},
  {Spangler}, {Stawarz}, {Steenkamp}, {Stegmann}, {Superina}, {Tam},
  {Tavernet}, {Terrier}, {van Eldik}, {Vasileiadis}, {Venter}, {Vialle},
  {Vincent}, {Vivier}, {V{\"o}lk}, {Volpe}, {Wagner}, {Ward}, {Zdziarski}, \&
  {Zech}}]{aharonian08}
{Aharonian}, F., {Akhperjanian}, A.~G., {Barres de Almeida}, U., {et~al.} 2008,
  \aap, 477, 353, \dodoi{10.1051/0004-6361:20078516}

\bibitem[{{Aharonian} \& {Atoyan}(1996)}]{aharonian_atoyan96}
{Aharonian}, F.~A., \& {Atoyan}, A.~M. 1996, \aap, 309, 917

\bibitem[{{Celli} {et~al.}(2020){Celli}, {Aharonian}, \& {Gabici}}]{Celli2020}
{Celli}, S., {Aharonian}, F., \& {Gabici}, S. 2020, \apj, 903, 61,
  \dodoi{10.3847/1538-4357/abb805}

\bibitem[{{Chernyakova} {et~al.}(2011){Chernyakova}, {Malyshev}, {Aharonian},
  {Crocker}, \& {Jones}}]{Chernyakova2011}
{Chernyakova}, M., {Malyshev}, D., {Aharonian}, F.~A., {Crocker}, R.~M., \&
  {Jones}, D.~I. 2011, \apj, 726, 60, \dodoi{10.1088/0004-637X/726/2/60}

\bibitem[{{Cox}(2005)}]{cox05}
{Cox}, D.~P. 2005, \araa, 43, 337,
  \dodoi{10.1146/annurev.astro.43.072103.150615}

\bibitem[{{Fujinaga} {et~al.}(2011){Fujinaga}, {Bamba}, {Dotani}, {Ozaki},
  {P{\"u}:Hlhofer}, {Wagner}, {Reimer}, {Funk}, \& {Hinton}}]{fujinaga11}
{Fujinaga}, T., {Bamba}, A., {Dotani}, T., {et~al.} 2011, \pasj, 63, S857,
  \dodoi{10.1093/pasj/63.sp3.S857}

\bibitem[{{Giacinti} {et~al.}(2018){Giacinti}, {Kachelrieẞ}, \&
  {Semikoz}}]{Giacinti}
{Giacinti}, G., {Kachelrieẞ}, M., \& {Semikoz}, D.~V. 2018, \jcap, 2018, 051,
  \dodoi{10.1088/1475-7516/2018/07/051}

\bibitem[{{Giunti} {et~al.}(2022){Giunti}, {Acero}, {Khelifi}, {Kosack},
  {Lemiere}, \& {Terrier}}]{xmm22}
{Giunti}, L., {Acero}, F., {Khelifi}, B., {et~al.} 2022, arXiv e-prints,
  arXiv:2209.09566.
\newblock \doarXiv{2209.09566}

\bibitem[{{Jansson} \& {Farrar}(2012)}]{jansson_farrar}
{Jansson}, R., \& {Farrar}, G.~R. 2012, \apj, 757, 14,
  \dodoi{10.1088/0004-637X/757/1/14}

\bibitem[{{Kafexhiu} {et~al.}(2014){Kafexhiu}, {Aharonian}, {Taylor}, \&
  {Vila}}]{kafexhiu14}
{Kafexhiu}, E., {Aharonian}, F., {Taylor}, A.~M., \& {Vila}, G.~S. 2014, \prd,
  90, 123014, \dodoi{10.1103/PhysRevD.90.123014}

\bibitem[{{Kelner} {et~al.}(2006){Kelner}, {Aharonian}, \&
  {Bugayov}}]{Kelner2006}
{Kelner}, S.~R., {Aharonian}, F.~A., \& {Bugayov}, V.~V. 2006, \prd, 74,
  034018, \dodoi{10.1103/PhysRevD.74.034018}

\bibitem[{{Koldobskiy} {et~al.}(2021){Koldobskiy}, {Kachelrie{\ss}},
  {Lskavyan}, {Neronov}, {Ostapchenko}, \& {Semikoz}}]{aafrag}
{Koldobskiy}, S., {Kachelrie{\ss}}, M., {Lskavyan}, A., {et~al.} 2021, arXiv
  e-prints, arXiv:2110.00496.
\newblock \doarXiv{2110.00496}

\bibitem[{{Lau} {et~al.}(2019){Lau}, {Rowell}, {Voisin}, {Blackwell}, {Burton},
  {Braiding}, {Wong}, {Fukui}, \& {Casanova}}]{Lau19}
{Lau}, J.~C., {Rowell}, G., {Voisin}, F., {et~al.} 2019, \mnras, 483, 3659,
  \dodoi{10.1093/mnras/sty3326}

\bibitem[{{Prosekin} {et~al.}(2015){Prosekin}, {Kelner}, \&
  {Aharonian}}]{prosekin15}
{Prosekin}, A.~Y., {Kelner}, S.~R., \& {Aharonian}, F.~A. 2015, \prd, 92,
  083003, \dodoi{10.1103/PhysRevD.92.083003}

\bibitem[{{Reichherzer} {et~al.}(2020){Reichherzer}, {Becker Tjus}, {Zweibel},
  {Merten}, \& {Pueschel}}]{Reichherzer2020}
{Reichherzer}, P., {Becker Tjus}, J., {Zweibel}, E.~G., {Merten}, L., \&
  {Pueschel}, M.~J. 2020, \mnras, 498, 5051, \dodoi{10.1093/mnras/staa2533}

\bibitem[{{Reichherzer} {et~al.}(2022){Reichherzer}, {Merten}, {D{\"o}rner},
  {Becker Tjus}, {Pueschel}, \& {Zweibel}}]{Reichherzer2022}
{Reichherzer}, P., {Merten}, L., {D{\"o}rner}, J., {et~al.} 2022, SN Applied
  Sciences, 4, 15, \dodoi{10.1007/s42452-021-04891-z}

\bibitem[{{Strong} {et~al.}(2007){Strong}, {Moskalenko}, \&
  {Ptuskin}}]{strong07}
{Strong}, A.~W., {Moskalenko}, I.~V., \& {Ptuskin}, V.~S. 2007, Annual Review
  of Nuclear and Particle Science, 57, 285,
  \dodoi{10.1146/annurev.nucl.57.090506.123011}

\bibitem[{{Vladimirov} {et~al.}(2012){Vladimirov}, {J{\'o}hannesson},
  {Moskalenko}, \& {Porter}}]{vladimirov12}
{Vladimirov}, A.~E., {J{\'o}hannesson}, G., {Moskalenko}, I.~V., \& {Porter},
  T.~A. 2012, \apj, 752, 68, \dodoi{10.1088/0004-637X/752/1/68}

\bibitem[{{Zabalza}(2015)}]{naima}
{Zabalza}, V. 2015, Proc.~of International Cosmic Ray Conference 2015, 922

\end{thebibliography}
\end{document}